\documentclass[aps,prd,twocolumn,amsmath,amssymb,amsfonts,nofootinbib,superscriptaddress,altaffilletter]{revtex4-1}

\usepackage{graphicx}
\usepackage{hyperref}
\hypersetup{
  bookmarksopen=true
}
\usepackage{ulem}
\usepackage{amssymb}
\usepackage{amsmath}
\usepackage{color}
\usepackage{supertabular}
\usepackage{dcolumn}

\def\fdot{f^{(1)}}

\newcommand{\SNR}{\textrm{SNR}}

\def\sci#1#2{#1\times10^{#2}}

\def\RAJ{\textrm{RA}_{\textrm J2000}}
\def\DECJ{\textrm{DEC}_{\textrm J2000}}

\begin{document}

\title{Results from high-frequency all-sky search for continuous gravitational waves from small-ellipticity sources}

\author{Vladimir Dergachev}
\email{vladimir.dergachev@aei.mpg.de}
\affiliation{Max Planck Institute for Gravitational Physics (Albert Einstein Institute), Callinstrasse 38, 30167 Hannover, Germany}
\affiliation{Leibniz Universit\"at Hannover, D-30167 Hannover, Germany}

\author{Maria Alessandra Papa}
\email{maria.alessandra.papa@aei.mpg.de}
\affiliation{Max Planck Institute for Gravitational Physics (Albert Einstein Institute), Callinstrasse 38, 30167 Hannover, Germany}
\affiliation{Leibniz Universit\"at Hannover, D-30167 Hannover, Germany}
\affiliation{University of Wisconsin Milwaukee, 3135 N Maryland Ave, Milwaukee, WI 53211, USA}

\begin{abstract}
We present the results of an all-sky search for continuous gravitational wave signals with frequencies in the 1700-2000\,Hz range from neutron stars with ellipticity of $\approx 10^{-8}$. The search employs the Falcon analysis pipeline \cite{Dergachev:2019wqa} on LIGO O2 public data. 
Our results improve by a factor greater than 5 over \cite{lvc_O2_allsky}. This is a huge leap forward: it takes an entirely new generation of gravitational wave detectors to achieve a 10-fold sensitivity increase over the previous generation \cite{Reitze:2019iox}. 
Within the probed frequency range and aside from the detected outliers, we can exclude neutron stars with ellipticity of $10^{-8}$ within 65\,pc of Earth. 
We set upper limits on the gravitational wave amplitude that hold even for worst-case signal parameters. New outliers are found, some of which we are unable to  associate with any instrumental cause. If any were associated with a rotating neutron star, this would likely be the fastest neutron star today.
\end{abstract}

\maketitle

Detectable continuous gravitational waves are expected from fast rotating neutron stars if they present some sort of non-axially symmetric deformation. The deformation in this context is usefully described by the {\it {ellipticity}} of the object, defined as $I_{zz}/ (I_{xx}-I_{yy})$, where $I$ is the moment of inertia tensor of the star and $\hat{z}$ is along the star's rotation axis \cite{Jaranowski:1998qm}. 

In our previous paper \cite{O2_falcon} we searched for gravitational wave emission in the 500-1700\,Hz range, targeting objects with ellipticity of $10^{-8}$. 
We found a number of outliers, some corresponding to known instrumental artifacts, some in data with pristine frequency spectrum. 

Ellipticities of $10^{-8}$ are interesting: they are well within the range of what the neutron star crusts can sustain \cite{Gittins:2020cvx} and observational indications exist that millisecond pulsars might have a {\it{minimum}} ellipticity of $\approx 10^{-9}$ \cite{ellipticity}.

The search \cite{O2_falcon} using the Falcon pipeline \cite{Dergachev:2019wqa, loosely_coherent, loosely_coherent2, loosely_coherent3} was performed on public LIGO data from the O2 science run. 
In order to investigate the consistency of our outliers waveforms with new data we would need at least a few Hz around the outliers frequency over the entire duration of a more sensitive data set. This data set is the O3 data, and we will be able to access it after it is released, in April 2021 \cite{DataManagementPlan}. 

The LIGO bulk data releases do not include any auxiliary-channel data \cite{O2_env}, thus we are limited to the analysis of the gravitational wave channel to identify detector artifacts.
Given the amount of contamination found in the O1 data, we would not be surprised if some of the outliers from \cite{O2_falcon} could be easily attributed to  detector disturbances.

So far none of our outliers from \cite{O2_falcon} have been linked to any instrumental cause, nor have results of follow-ups on O3 data been reported. This 
leaves us in the same state of knowledge as when \cite{O2_falcon} came out and justifies further expanding the search range to higher 
frequencies where fewer detector artifacts are commonly observed.

\begin{figure}[htbp]
\includegraphics[width=3.3in]{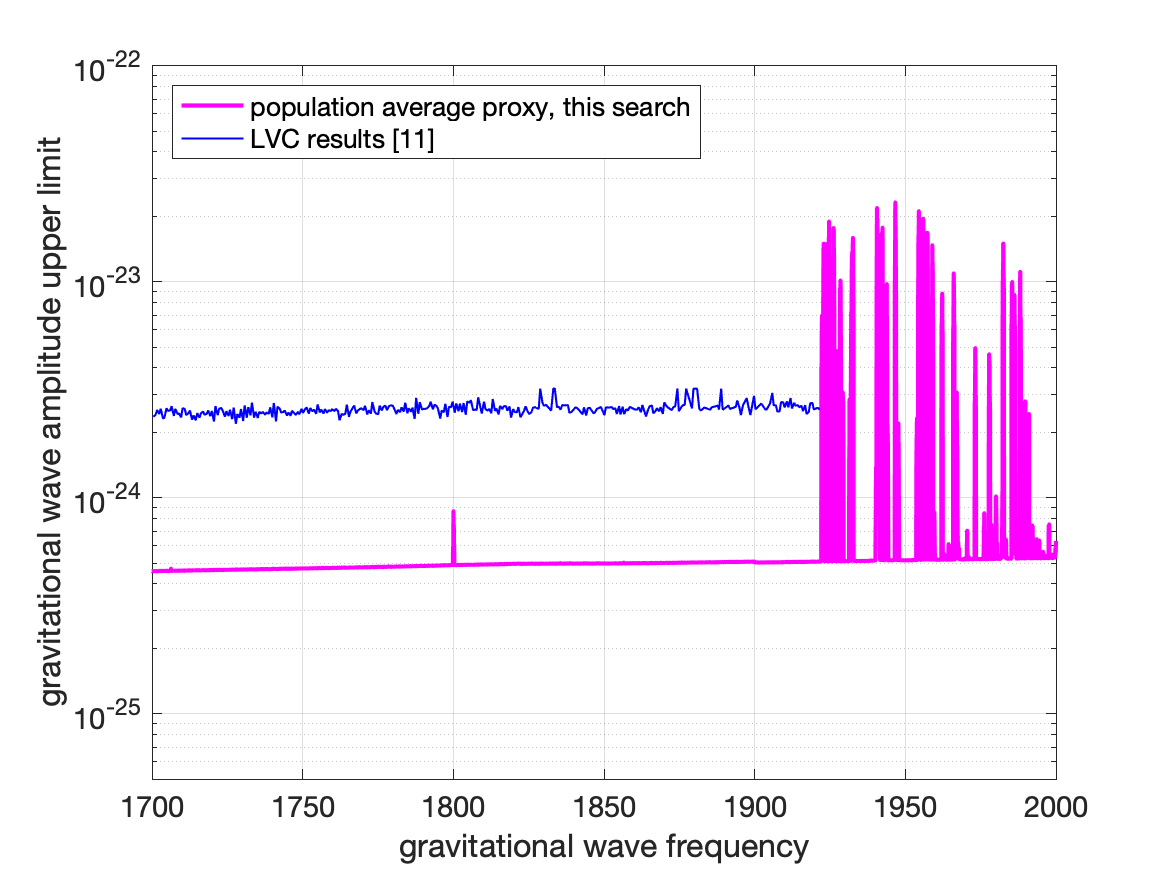}
\caption[Spindown range]{
\label{fig:amplitudeULs}
Gravitational wave intrinsic amplitude $h_0$ upper limits at 95\% confidence as a function of signal frequency. The upper limits are a measure of the sensitivity of the search. We compare with the latest LIGO/Virgo all-sky results in this frequency range \cite{lvc_O2_allsky}, which are a factor $\gtrsim$ 5 less constraining than ours. The results in \cite{Palomba:2019vxe} are a factor of 2 less constraining.
}
\end{figure}

\begin{table}[htbp]
\begin{center}
\begin{tabular}{D{.}{.}{3}D{.}{.}{5}D{.}{.}{4}D{.}{.}{4}D{.}{.}{4}l}\hline
\multicolumn{1}{c}{SNR}   &  \multicolumn{1}{c}{Frequency} & \multicolumn{1}{c}{Spindown} &  \multicolumn{1}{c}{$\RAJ$}  & \multicolumn{1}{c}{$\DECJ$} & Comment \\
\multicolumn{1}{c}{}	&  \multicolumn{1}{c}{Hz}	&  \multicolumn{1}{c}{pHz/s} & \multicolumn{1}{c}{degrees} & \multicolumn{1}{c}{degrees} & \\
\hline \hline
% snr & % frequency & % spindown & % ra & % dec & % Description
1025.8 & 1991.09236 & -1.08 & 300.802 & -14.320 & ip14\\
19.6 & 1991.12667 & -1.42 & 300.898 & 11.804 & ip14 echo\\
17.7 & 1998.75364 & -7.86 & 238.873 & 84.926 & H1 line\\
17.4 & 1891.75674 & -8.22 & 171.140 & 57.628 & \\
16.1 & 1892.99106 & -1.08 & 216.530 & -46.769 & \\
16.1 & 1985.75899 & 3.22 & 198.326 & -73.539 & L1 line\\

\hline
\end{tabular}
\caption[Outliers produced by the detection pipeline]{Outliers produced by the detection pipeline. Only the highest-SNR outlier is shown for each 0.1\,Hz frequency region. Outliers marked ``ip14'' are due to a simulated signal ``hardware-injected'' during the science run for validation purposes. Its parameters are listed in Table \ref{tab:injections}. Outliers marked with ``line'' have strong narrowband disturbances near the outlier frequency.
Signal frequencies refer to GPS epoch $1183375935$.} 
\label{tab:Outliers}
\end{center}
\end{table}

\begin{table}[htbp]
\begin{center}
\begin{tabular}{lD{.}{.}{6}cD{.}{.}{5}D{.}{.}{4}}
\hline
Label & \multicolumn{1}{c}{Frequency} & \multicolumn{1}{c}{Spindown} & \multicolumn{1}{c}{$\RAJ$} & \multicolumn{1}{c}{$\DECJ$} \\
 & \multicolumn{1}{c}{Hz} & \multicolumn{1}{c}{pHz/s} & \multicolumn{1}{c}{degrees} & \multicolumn{1}{c}{degrees} \\
\hline \hline
ip14  &  1991.092400 &  -1  & 300.80284  & -14.32394 \\
\hline
\end{tabular}
\caption[Parameters of hardware injections]{Parameters of the hardware-injected simulated continuous wave signals during the O2 data run (GPS epoch $1130529362$).}
\label{tab:injections}
\end{center}
\end{table}

In the new  $1700-2000$\,Hz band we expand the first-order frequency derivative search range to $  \sci{-7.5}{-12} \leq f_1  \leq \sci{2.5}{-12}$\,Hz/s, consistent with our ellipticity target at a level $\lesssim 10^{-8}$. We leave unchanged the rest of the search parameters. 

Our search looks for signals that are phase coherent over a certain time-span, called the coherence-length, and that evolve slowly over longer timescales. The search detects with highest efficiency continuous wave signals from isolated objects, with constant amplitude and frequency evolution that can be described by a Taylor expansion around nominal frequency and frequency derivative values, see for example Section II of \cite{Jaranowski:1998qm}. We will refer to these signals as the ``{ {IT2 model}}'' continuous wave signals. Here ``I'' indicates the source being isolated and ``T2'' indicates an intrinsic frequency evolution given by a Taylor polynomial of degree 2. In the literature IT2 signals are used as a reference to measure search performance, and for upper limits.

This search employs a first coherence-length of 12 hours. Regions of parameter space associated with high SNR results are searched again with a coherence length of 24 hours. The process is repeated with a coherence length of 48 and finally of 144 hours, as done in \cite{O2_falcon}. This search is considerably more computationally intensive per unit frequency interval compared to \cite{O2_falcon}, and in fact it took approximately the same amount of compute cycles to cover 300 Hz in this frequency range than it did to cover 1200 Hz at the lower frequencies investigated in \cite{O2_falcon}. 

High-frequency searches are not only interesting because they can probe lower neutron star deformations. Detecting continuous gravitational waves between 1700 and 2000 Hz  would also unveil the fastest rotating neutron star and would be a breakthrough in explaining why electro-magnetic observations have not revealed any pulsar rotating faster than about 700 Hz, while the maximum rotation frequency cannot be lower than 1.2 kHz \cite{Haskell:2018nlh}.

\begin{figure}[htbp]
\includegraphics[width=3.3in]{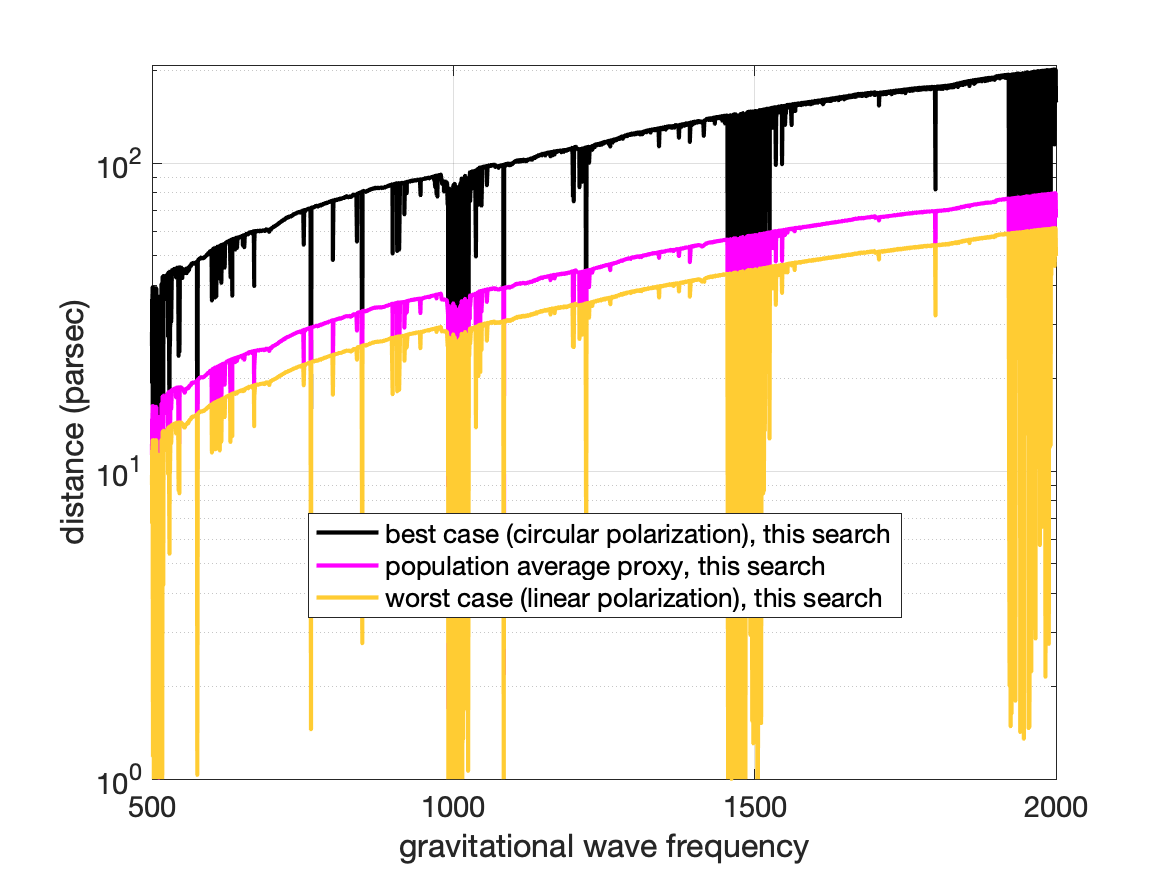}%{O2_full_ul_distance}
\caption[Spindown range]{
\label{fig:distance}
Reach of the search for stars with ellipticity of $10^{-8}$, including the results from previous paper \cite{O2_falcon} below 1700 Hz. The X axis is the gravitational wave frequency, which is twice the pulsar rotation frequency for emission due to an equatorial ellipticity. R-modes and other emission mechanisms give rise to emission at different frequencies. The top curve (black) shows the results for a population of circularly polarized sources; The middle curve (magenta) holds for a population of sources with random orientations; The bottom curve (yellow) to linearly polarized sources. }
\end{figure}

The new search could see such low ellipticity objects up to 200\,pc away (Figure \ref{fig:distance}) for circularly polarized signals, and we would expect to detect an arbitrary oriented source within 65\,pc.  The gravitational wave amplitude upper limits from this search are shown in Figure \ref{fig:amplitudeULs} and are available in numerical form in \cite{data}. 

Our upper limits can be translated into limits on gravitational waves from  boson condensates around black holes \cite{boson1,boson2}, which are expected to emit monochromatic continuous wave signals \cite{discovering_axions}. 
We leave it to the interested reader to constrain from our upper limits physical quantities of interest, based on the specific model they wish to consider. Assuming the ensemble signal of \cite{Zhu:2020tht} from a galactic population of $O(10^8)$ isolated stellar mass black holes with maximum mass 30 $M_\odot$ and maximum initial spin uniformly distributed in  [0,1], our results extend the boson mass exclusion region to $4.0 \times 10^{-12}$ eV (Fig. 18 of \cite{Zhu:2020tht}).

A key result of our search are several outliers (Table~\ref{tab:Outliers}), two of which are located in bands with clean frequency spectrum. 

\begin{figure}[htbp]
\includegraphics[width=3.3in]{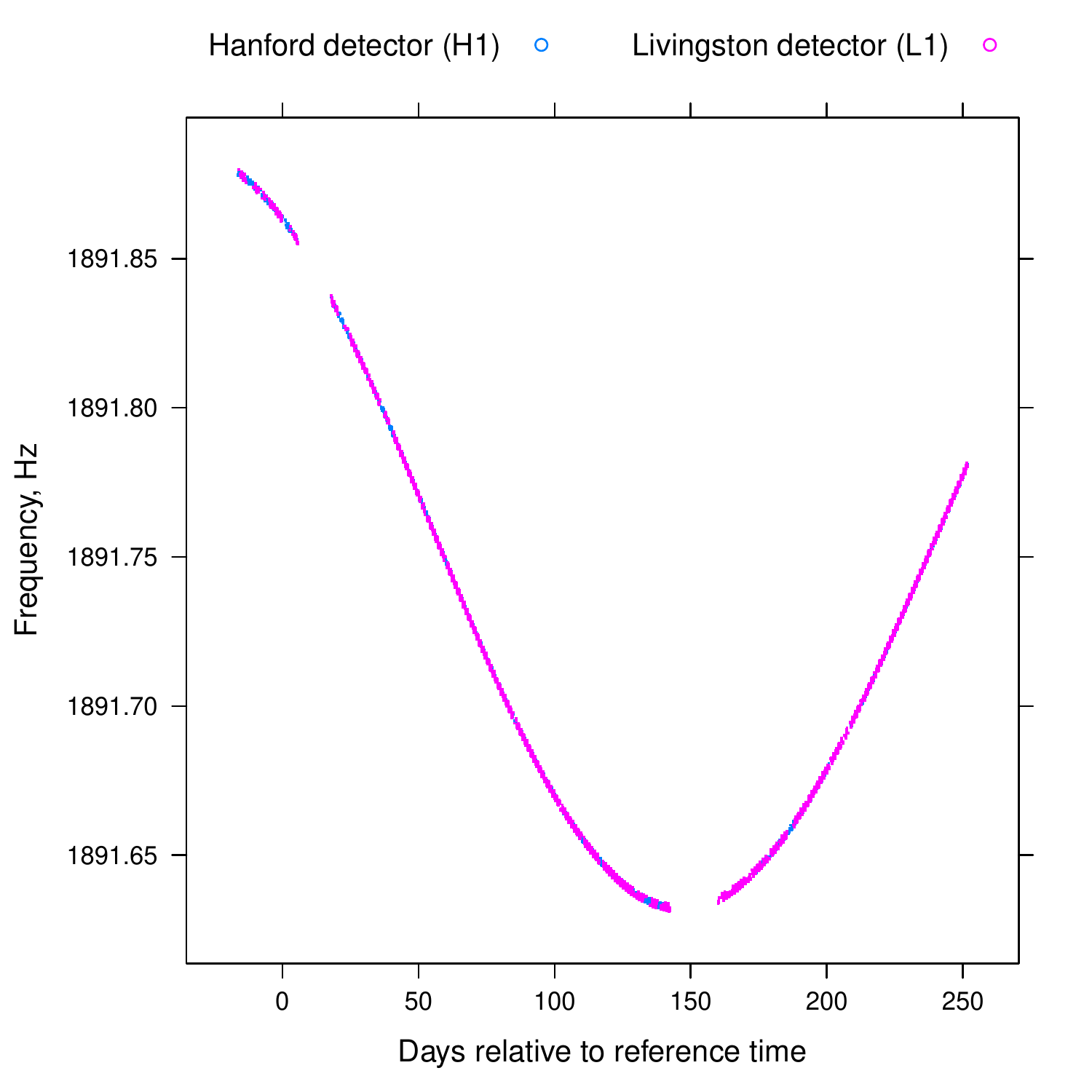}
\caption[Outlier map]{
\label{fig:outlier_evolution}
Apparent frequency of a signal with parameters equal to those of the outlier at 1891.76\,Hz, at the detectors.  The difference in Doppler shifts between interferometers is small compared to the Doppler shifts from the Earth's orbital motion. We recall that the reference time is at GPS epoch $1183375935$.
}
\end{figure}
\begin{figure}[htbp]
\includegraphics[width=3.3in]{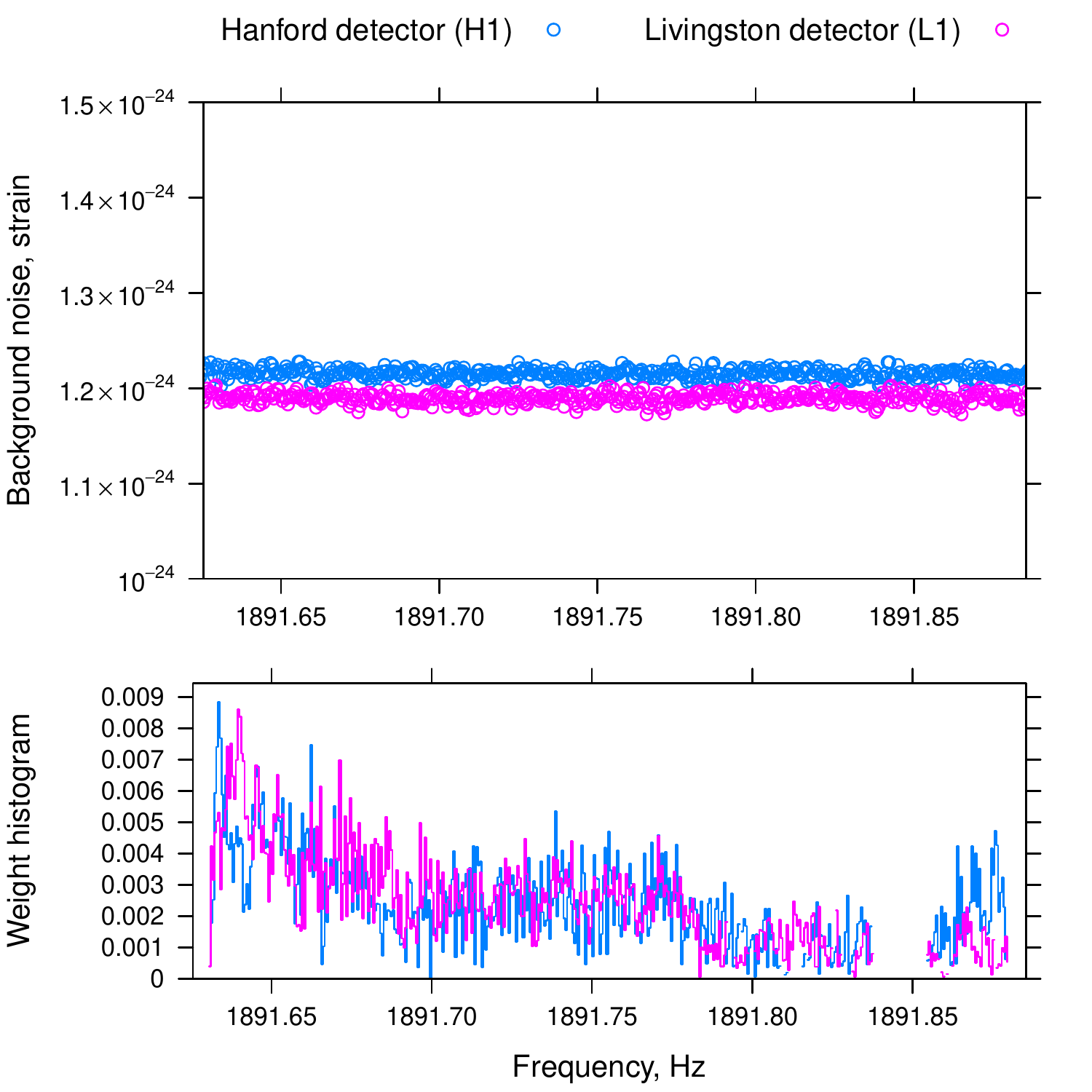}
\caption[Outlier spectrum]{
\label{fig:outlier_hist}
The top plot shows the average amplitude spectral density around the frequency of the outlier at 1891.76\,Hz. The bottom plot shows how much the power in each frequency bin would contribute, over the course of the O2 run, to the total power estimated by the Falcon pipeline for a signal with the parameters of the outlier at $\approx$ 1891.8 Hz. High-weight values correspond to bins with greater contribution and the sum of all the weights is 1 for each curve. The gap corresponds to a break in the O2 run and matches the gap in the frequency evolution plot \ref{fig:outlier_evolution}.
}
\end{figure}
The outlier with the highest signal-to-noise ratio (SNR) is at 1891.76\,Hz and it is not associated with any reported detector artifact \cite{Covas:2018oik}. At such high frequency the Doppler shift changes the frequency of the received signal by 0.25\,Hz during O2 (Figure \ref{fig:outlier_evolution}). The spectrum in this frequency band is clean (Figure \ref{fig:outlier_hist}) with no evidence of any contamination in either detector. 

We examine the SNR build-up for the two highest-SNR unmarked outliers. An IT2 continuous wave signal would be expected to have $\approx$ constant rate of signal accumulation and be consistently identified by other search methods. Our two outliers deviate from this behaviour, showing SNR accumulation peaks in subsets of the data (see appendix for more details and cautions on over-interpreting this fact). A semi-coherent matched-filter search of the type used in \cite{EHO2}, combining the results of 29 coherent searches, each on stretches of data 126 hours long, does not return particularly significant results. 

In spite of significant search efforts, continuous gravitational waves have yet to be detected. As we have pointed out in \cite{O2_falcon} the reason could be that the signal strength is lower than what searches can nowadays detect. But it could also be that the signal may not be an IT2 signal. There are many reasons for which this could happen: the signal itself might not be perfectly phase coherent. After all, we do know that pulsars exist which glitch and that present spin irregularities. The pulse profiles of pulsars show that the physics of these objects is rich and complex. It may also be that the signal is perfectly regular but different than the IT2  model: For instance the star could have a companion or its spin axis could be slowly precessing. 

The loosely coherent search employed by the Falcon pipeline is specifically designed to be more robust than matched-filter based searches with respect to deviations of the signal from the IT2 model. Signals can be imagined that would produce the observed results but these are a-posteriori considerations that do not add to the significance of the present outliers. The reason why we highlight these outliers is that it is time to entertain the possibility that continuous waves may not come in the IT2 signal form. The detection of such signals will be harder and it begins with the careful consideration of all outliers that cannot be reasonably ascribed to instrumental causes. Such are the outliers presented here and in \cite{O2_falcon}. The big sensitivity improvement of this search over previous ones makes these results uniquely interesting. The small parameter range associated with the outliers identified here, enables dedicated follow-ups on new data that will advance our understanding of these first observations.

The number of known spectral instrumental artifacts typically decreases with frequency, many of them being harmonics of low-frequency noise sources and the higher frequency ``violin'' modes being carefully segregated in specific frequency bands to avoid contamination of the rest of the spectrum.
Thus the fact that we see two outliers in a 300\,Hz band makes them interesting. If they are due to a noise source this provides a unique opportunity to identify it and remove it, perhaps unveiling a new class of ``weak" contaminants. If no instrumental noise source is found, these signals should be a focus of thorough future investigations.

\begin{acknowledgments}
The authors thank the scientists, engineers and technicians of LIGO, whose hard work and dedication produced the data that made this search possible.

The search was performed on the ATLAS cluster at AEI Hannover. We thank Bruce Allen, Carsten Aulbert and Henning Fehrmann for their support. We also thank Bernard Schutz and Bruce Allen for interesting and helpful discussions on the significance of the results presented here.

This research has made use of data, software and/or web tools obtained from the LIGO Open Science Center (\url{https://losc.ligo.org}), a service of LIGO Laboratory, the LIGO Scientific Collaboration and the Virgo Collaboration.  LIGO is funded by the U.S. National Science Foundation. Virgo is funded by the French Centre National de Recherche Scientifique (CNRS), the Italian Istituto Nazionale della Fisica Nucleare (INFN) and the Dutch Nikhef, with contributions by Polish and Hungarian institutes.
\end{acknowledgments}

\appendix*

\section{SNR accumulation}
\label{snr_accumulation}

For an ideal signal, in ideal noise and ideal data any reasonable measure of SNR should display a steady increase as more data is used. For an instrumental or environmental disturbance that does not match the signal waveform, it is unlikely that one would observe the same steady accumulation of SNR.  One could then think of using this SNR-accumulation ``signature"  to discriminate between candidates that stem from signals and ones that stem from disturbances. The purpose of this section is to caution the reader against over-interpreting SNR-accumulation curves, especially when they come from just-above-threshold candidates.

In the real world there are a number of factors that produce deviations of the SNR-accumulation with respect to the ideal case, even for an ideal signal: 
the noise of the instruments is not constant, so the SNR-accumulation is different than if the detectors were perfectly stable. The extreme manifestation of this is when there is no science-quality data available from one or either of the detectors for a period of time. Since the data from both detectors is coherently combined, this impacts the SNR-accumulation. 

The sensitivity variations of the detectors are not completely random. For example there are evident day/night modulations, from anthropogenic activity. Since the beam-pattern functions of the instruments are not constant in time for a given source, unfortunate combinations may occur of the higher sensitivity periods being affected by higher noise.  This again impacts the SNR-accumulation.

Of course there are time scales of the search set-up and of the disturbances for which some of the effects average out, but fundamentally the SNR-accumulation is the projection of a complex process affected by a number of factors, on a single figure of merit. To draw meaningful conclusions about any signal candidate based on the associated SNR-accumulation is not trivial. 

One of the ways that one could characterise the SNR-accumulation is with extensive Monte Carlos, using fake signals on the real data. Our past attempts to find a safe {\it{and}} effective veto for just-above threshold candidates, based on SNR-accumulation for all-sky searches, have not been successful, and we are not aware of any SNR-accumulation veto being systematically incorporated in any all-sky search.  

SNR-accumulation becomes more reliable as the signal strength increases, ``washing-out" all other effects. But the loud-signal regime is not particularly interesting in the O2 data because we do not expect to observe very loud signals: after all, no continuous wave signals have been detected in any data collected earlier. For this reason we have not revisited the question of setting up a veto based on SNR-accumulation for this search.  

The Falcon pipeline stores SNR-accumulation data for all outliers, for historical reasons (from past attempts to construct vetoes) and to highlight whether very egregious deviations exist, that sometimes point to very evident disturbances that can be easily identified in the strain data. 

While this was not the case for the two outliers of Table I, their behaviour was also not what is expected for loud signals. For the interested reader, in this section we show the SNR-accumulation plots for the outliers. We think that it is though necessary to provide some sense of the broader context for these and so we also show SNR-accumulation plots for fake signals.
\begin{figure}[htbp]
\includegraphics[width=3.3in]{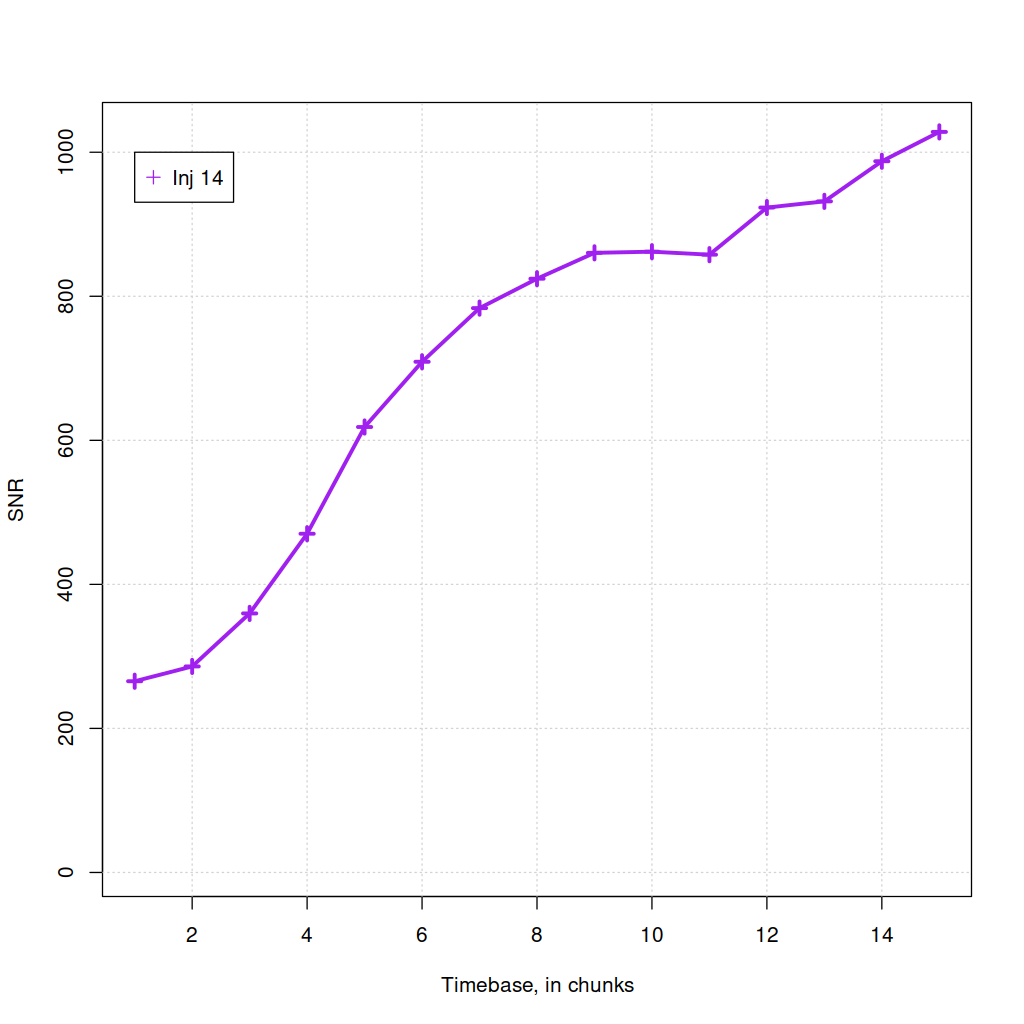}
\caption[SNR-accumulation for hardware injection 14]{
\label{fig:snr14}
SNR versus search duration for the search results associated with hardware injection signal ip14. Because the signal is very loud we see a mostly monotonic accumulation of SNR, albeit with deviations from the expected ideal behaviour due to the non-stationarity of the data.  
}
\end{figure}

We partition the O2 data into 15 chunks of equal duration; we perform searches like the last stage of our search with different durations. Specifically we consider 15 durations, for each using the first $N_i \in \left\{1, \ldots, 15\right\}$ chunks. 

\begin{figure}[htbp]
\includegraphics[width=3.3in]{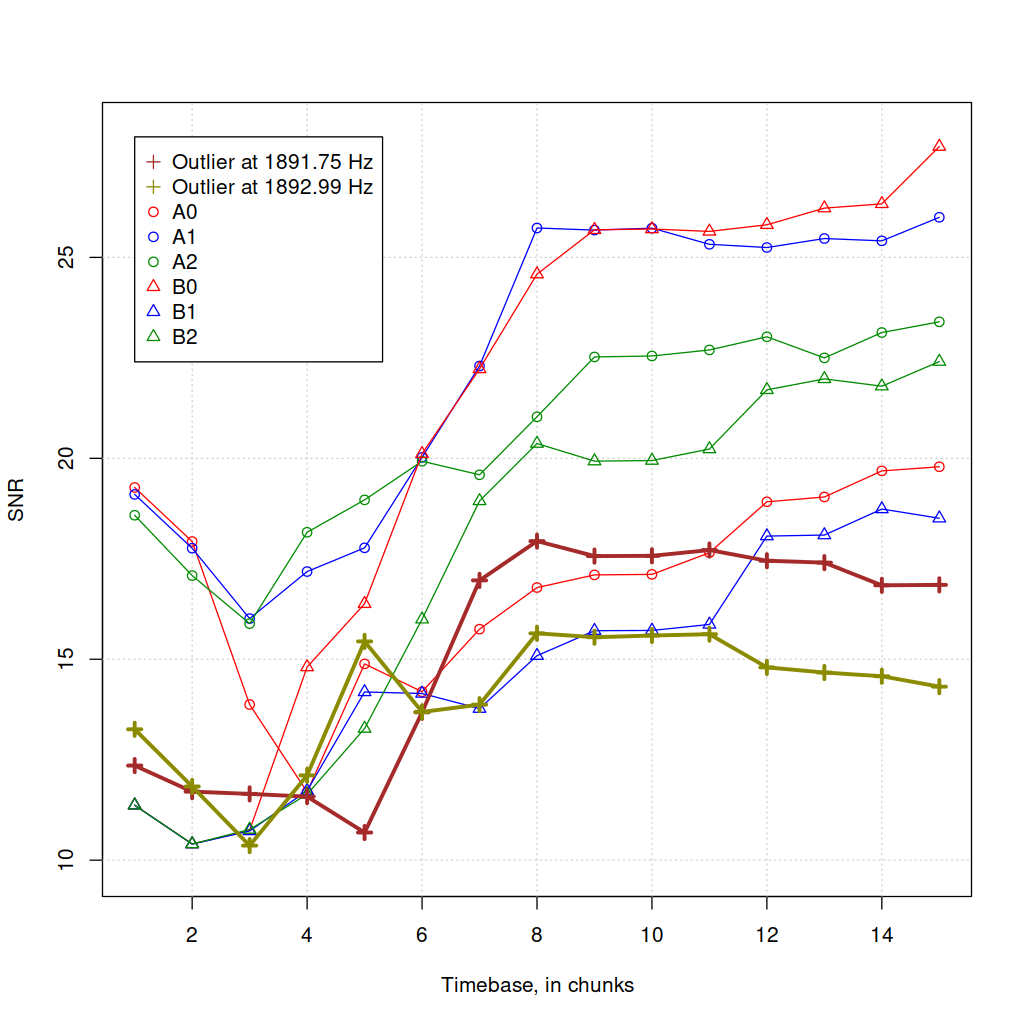}
\caption[SNR-accumulation for small signals]{
\label{fig:snr}
SNR versus search duration for the outliers from the search, at 1891.75\,Hz and at at 1892.99\,Hz (thick curves, marked by crosses) and for fake signals (see text)
}
\end{figure}

Figure \ref{fig:snr14} shows the SNR-accumulation for the hardware injected signal ip14. This is a simulated pulsar signal introduced into the strain channel by artificially moving the mirrors. With an intrinsic strain amplitude of $3.4 \times 10^{-24}$ this is a rather loud signal for our search, reaching SNR of $1025$ when using all the data. For reference our population-average upper limit at that frequency is $\approx 5.5 \times 10^{-25}$. The LVC search \cite{lvc_O2_allsky} does not extend beyond 1922 Hz, but extrapolating the upper limit at that frequency at $\approx 2.6 \times 10^{-24}$, we conclude that this signal would be detectable by that search as well. 

We see that the SNR increases with increasing data, but does not follow a square-root law, as expected in the ideal case. It is instructive to appreciate how much the non-stationarities of the real situation impact the SNR-accumulation, even for a very loud signal. 

Figure \ref{fig:snr} shows the SNR-accumulation plot of the outliers at 1891.75\,Hz and 1892.99\,Hz, together with SNR-accumulation for 6 fake signals.
The fake signals have a frequency that is close to that of the outliers, so that the character of the noise is not completely different, but far away enough that the search results are completely independent: 0.5 Hz lower than the lowest-frequency outlier, i.e. at 1891.25\,Hz. 

The intrinsic strain amplitude of the fake signal is set to the worst-case upper limit in that frequency range. The sky-position value is set equal to that of the 1891.75\,Hz outlier's. We consider three spin-down values : $\fdot_k= (\fdot_0 + k \times 10^{-13})$\,Hz/s, where $\fdot_0$ is the spin-down value associated with the 1891.75\,Hz outlier and $10^{-13}$\,Hz/s is $1/10$-th of the uncertainty in spin-down for outlier signals \cite{O2_falcon,Dergachev:2019wqa}. We consider signals with different initial phases, namely the ``A'' series with phase set to $0$ at the reference time, and the ``B'' series with phase set to $\pi$. 

We observe that there is large variability in SNR depending on the signal. The outliers' curves display smaller SNRs than the fake signals, which is to be expected, because the 95\% confidence worst-case upper limit strain is larger than most detectable signal strains. 

\begin{figure}[htbp]
\includegraphics[width=3.3in]{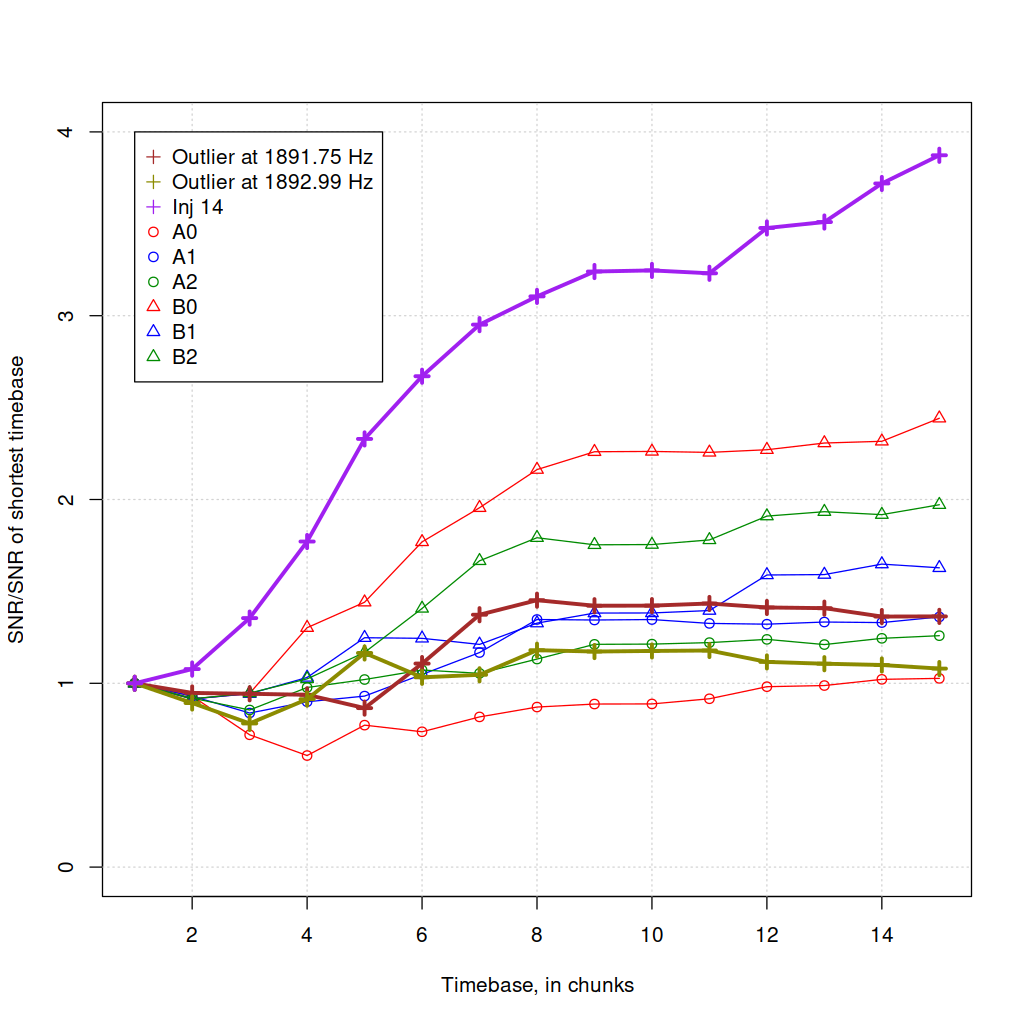}
\caption[Relative SNR-accumulation for small signals]{
\label{fig:rel_snr}
Relative SNR (\SNR(N$_i$)/\SNR(N$_1$)) versus search duration for all signals shown in figures \ref{fig:snr14} and \ref{fig:snr}. The SNR values are divided by the SNR of the first (shortest) chunk.
}
\end{figure}

We compare the SNR-accumulation of all signals and outliers independently of their SNR value by dividing the SNR for each chunk by the SNR of the shortest chunk (the first chunk). This relative SNR is shown in figure \ref{fig:rel_snr}. 
The loud hardware injection gains a factor of $\approx 4$ in SNR, consistent with timebase increase by a factor of 15. 
The software fake signals show smaller overall increase and this illustrates how much random fluctuations as well as the non-stationarities of the problem may impact the resulting SNR. The non-linear interaction between the analysis pipeline and the underlying noise is very hard to model for small signals. 

In conclusion, the SNR-accumulation curve can be of help for very loud signals, and is a diagnostic tool that may help to identify egregious disturbances. Based on extensive studies we have in the past found that it is hard to find a safe and reliable veto based on such plots for candidates with SNR of $\approx 20$ and smaller. The small-scale studies reported here, focused around the outliers' parameters, support that conclusion.


\newpage
\begin{thebibliography}{99}

\def\etal{{\it et al.}}

\bibitem{Jaranowski:1998qm}
P.~Jaranowski, A.~Krolak and B.~F.~Schutz,
``Data analysis of gravitational - wave signals from spinning neutron stars. 1. The Signal and its detection,''
Phys. Rev. D \textbf{58} (1998), 063001
doi:10.1103/PhysRevD.58.063001
[arXiv:gr-qc/9804014 [gr-qc]].


\bibitem{O2_falcon}
V.~Dergachev, M.~A.~Papa, 
``Results from the first all-sky search for continuous gravitational waves from small-ellipticity sources", 
Phys. Rev. Lett. \textbf{125}, no.17, 171101  (2020)
doi:10.1103/PhysRevLett.125.171101
[arXiv:2004.08334 [gr-qc]].


\bibitem{Gittins:2020cvx}
F.~Gittins, N.~Andersson and D.~I.~Jones,
``Modelling neutron star mountains,''
[arXiv:2009.12794 [astro-ph.HE]].

\bibitem{ellipticity}
G.~Woan, M.~D.~Pitkin, B.~Haskell, D.~I.~Jones, and P.~D.~Lasky, Evidence for a Minimum Ellipticity in Millisecond Pulsars, ApJL {\bf{863}} L40	(2018)

\bibitem{Dergachev:2019wqa}
V.~Dergachev and M.~A.~Papa,
``Sensitivity improvements in the search for periodic gravitational waves using O1 LIGO data,''
Phys. Rev. Lett. \textbf{123}, no.10, 101101 (2019)
doi:10.1103/PhysRevLett.123.101101
[arXiv:1902.05530 [gr-qc]].



\bibitem{loosely_coherent} 
V.~Dergachev, On blind searches for noise dominated signals: a loosely coherent approach, Class.\ Quantum Grav.\ {\bf 27}, 205017 (2010).

\bibitem{loosely_coherent2} 
V.~Dergachev, Loosely coherent searches for sets of well-modeled signals,
Phys.\ Rev.\ D {\bf 85}, 062003 (2012)  

\bibitem{loosely_coherent3} 
V.~Dergachev, Loosely coherent searches for medium scale coherence lengths, arXiv:1807.02351   


\bibitem{DataManagementPlan} LIGO Data Management Plan {texttt {https://dcc.ligo.org/public/0009/M1000066/025/LIGO-M1000066-v25.pdf}} (2017)


\bibitem{O2_env}
P.~Nguyen, R.~M.~S.~Schofield, A.~Effler, C.~Austin, V.~Adya, \etal, Environmental Noise in Advanced LIGO Detectors, arXiv:2101.09935 [astro-ph.IM]

\bibitem{lvc_O2_allsky}
B.~P.~Abbott \etal\ (LIGO Scientific Collaboration and Virgo Collaboration), All-sky search for continuous gravitational waves from isolated neutron stars using Advanced LIGO O2 data, Phys.\ Rev.\ D {\bf 100} 024004 (2019).

\bibitem{Reitze:2019iox}
D.~Reitze, R.~X.~Adhikari, S.~Ballmer, B.~Barish, L.~Barsotti, G.~Billingsley, D.~A.~Brown, Y.~Chen, D.~Coyne and R.~Eisenstein, \textit{et al.}
``Cosmic Explorer: The U.S. Contribution to Gravitational-Wave Astronomy beyond LIGO,''
Bull. Am. Astron. Soc. \textbf{51}, 035
[arXiv:1907.04833 [astro-ph.IM]].

\bibitem{Palomba:2019vxe} 
   C.~Palomba {\it et al.}, Direct constraints on ultra-light boson mass from searches for continuous gravitational waves, Phys. Rev. Lett. {\bf 123}, 171101 (2019)



\bibitem{Haskell:2018nlh}
B.~Haskell, J.~L.~Zdunik, M.~Fortin, M.~Bejger, R.~Wijnands and A.~Patruno,
``Fundamental physics and the absence of sub-millisecond pulsars,''
Astron. Astrophys. \textbf{620} (2018), A69
doi:10.1051/0004-6361/201833521

\bibitem{data} See EPAPS Document No. [number will be inserted by
publisher] for numerical values of upper limits, outlier tables and hardware injection parameters. Also at \url{https://www.aei.mpg.de/continuouswaves/O2Falcon500-1700}



\bibitem{boson1}
M.~Baryakhtar, R.~Lasenby, M.~Teo,  Black Hole Superradiance Signatures of Ultralight Vectors, Phys.\ Rev.\ D {\bf 96}, 035006s (2017)

\bibitem{boson2} 
A.~Arvanitaki, M.~Baryakhtar, R.~Lasenby, S.~Dimopoulos, S.~Dubovsky, Black Hole Mergers and the QCD Axion at Advanced LIGO, Phys.\ Rev.\ D {\bf 95}, 043001 (2017)

\bibitem{discovering_axions} A.~Arvanitaki, M.~Baryakhtar, and X.~Huang, Discovering the QCD axion with black holes and gravitational waves, 
Phys.\ Rev.\ D {\bf 91}, 084011 (2015)

\bibitem{Zhu:2020tht}
S.~J.~Zhu, M.~Baryakhtar, M.~A.~Papa, D.~Tsuna, N.~Kawanaka and H.~Eggenstein,
Characterizing the continuous gravitational-wave signal from boson clouds around Galactic isolated black holes, Phys.\ Rev.\ D {\bf 102}, 063020 (2020)


\bibitem{Covas:2018oik} 
P.~B.~Covas, A.~Effler, E.~Goetz, P.~M.~Meyers, A.~Neunzert, M.~Oliver, B.~L.~Pearlstone, V.~J.~Roma, R.~M.~S.~Schofield, \etal,
Identification and mitigation of narrow spectral artifacts that degrade searches for persistent gravitational waves in the first two observing runs of Advanced LIGO, Phys. Rev. D {\bf 97} 082002 (2018)


\bibitem{EHO2}
B.~Steltner, M.~A.~Papa, H.-B.~Eggenstein, B.~Allen, V.~Dergachev, R.~Prix, B.~Machenschalk, S.~Walsh, S.~J.~Zhu, S.~Kwang,
Einstein@Home all-sky search for continuous gravitational waves in LIGO O2 public data, arXiv:2009.12260


\end{thebibliography}
\end{document}